\documentstyle[prd,aps,epsfig,preprint]{revtex}
\tighten
\begin{document}
\thispagestyle{empty}
\baselineskip24pt
\draft
\begin{center}
{\large\bf Upon symmetry of baryon magnetic moments in the chiral and
the quark-soliton models}
\end{center}
 \begin{center}
{Elena~N.~Bukina$^a$, Vladimir~M.~Dubovik$^a$ and
Valery~S.~Zamiralov$^{b}$}
\hbox{\it $^a$Joint Institute for Nuclear Research,
          141980 Dubna, Moscow region, Russia} 
\hbox{\it $^b$ D.V. Skobeltsyn Institute of Nuclear Physics,
Moscow State University, Moscow, Russia}
\end{center}
\date{\today}
\begin{abstract}
It is shown that the baryon magnetic moment descriptions in the
frameworks of the chiral  model ChPT and the quark soliton $\chi QSM $
one are practically identical. The main difference from the 
traditional unitary
symmetry models proves to be in terms due to the pionic current 
contribution into the magnetic moments of nucleons and
in the prediction for the $\Lambda$
magnetic moment.
\end{abstract}

\section*{Introduction}

Since appearance of a theory of unitary symmetry and then of a quark model,
a problem of baryon octet magnetic moment description has been attracting many
theoretical efforts. It is well known that a traditional unitary symmetry
model with  only two parameters \cite{Gl} describes experimental data at 
the qualitative level.
The quark models (cf.,e.g.,\cite{Morp}) usually introduce the number of
parameters $\ge 3$ and are able to describe baryon magnetic moments
quantitatively better than to $10\%$. In general, these quark models have
strongly improved the agreement with experimental data.
However, at the present time experimental data have achieved
a level of $1 \% $ accuracy \cite{Mont}, which force theoreticians to
upgrade precision of a theoretical and phenomenological
description of the baryon magnetic moments. Many interesting models 
have been developed in order to solve this long-stayed problem from
various points of view, and we are able here to cite only
some of them [4-20].

Recently, baryon magnetic moments have been analyzed within two
independent chiral models \cite{Chang} and \cite{Kim}. It has been shown
there that the baryon magnetic moment symmetry
is based on unitary symmetry, and the chiral model ChPT \cite{Chang}
and quark soliton model \cite{Kim} just
describe the way in which this symmetry is broken.
We would try to establish a relation between these two models.

Besides, we shall show that a phenomenological model generated by the
unitary symmetry approach can be formulated in terms
of the electromagnetic baryon current, which proves to be quite  
close to these two models.

\section{Chiral model ChPT for the baryon magnetic moments}

Let us briefly remind  the ChPT model \cite{Chang} for the baryon octet
magnetic moments.It has been assumed that the leading $SU(3)_{f}$
breaking corrections to the magnetic moments have the same chiral
transformation properties as the strange quark mass operator and
the corresponding coefficients are of the order $m_{s}/\Lambda_{\chi}$,
$m_{s}$ and $\Lambda_{\chi}$, being strange quark mass and chiral
symmetry breaking scale, respectively.
The expressions for the baryon octet
magnetic moments in this model read \cite{Chang}

\begin{eqnarray}
\mu(p)=\frac{1}{3}(b_{1}+\alpha_{4})+(b_{2}+\alpha_{2})+\alpha_{1}+
\frac{1}{3}\alpha_{3}-\frac{1}{3}\beta_{1}
\nonumber\\
\mu(n)=-\frac{2}{3}(b_{1}+\alpha_{4})-\frac{2}{3}\alpha_{3}-
\frac{1}{3}\beta_{1}
\nonumber\\
\mu(\Sigma^{+})=\frac{1}{3}(b_{1}+\alpha_{4})+(b_{2}+\alpha_{2})-\alpha_{2}-
\frac{1}{3}\alpha_{4}-\frac{1}{3}\beta_{1}
\nonumber\\
\mu(\Sigma^{-})=\frac{1}{3}(b_{1}+\alpha_{4})-(b_{2}+\alpha_{2})+\alpha_{2}-
\frac{1}{3}\alpha_{4}-\frac{1}{3}\beta_{1}
\nonumber\\
\mu(\Xi^{0})=-\frac{2}{3}(b_{1}+\alpha_{4})+\frac{2}{3}\alpha_{3}-
\frac{1}{3}\beta_{1}
\nonumber\\
\mu(\Xi^{-})=\frac{1}{3}(b_{1}+\alpha_{4})-(b_{2}+\alpha_{2})+\alpha_{1}-
\frac{1}{3}\alpha_{3}-\frac{1}{3}\beta_{1}
\nonumber\\
\mu(\Lambda^{0})=-\frac{1}{3}(b_{1}+\alpha_{4})-
\frac{5}{9}\alpha_{4}-\frac{1}{3}\beta_{1}
\label{chang}
\end{eqnarray}

These expressions can easily be rewritten in the form demonstrating
that the chiral model ChPT \cite{Chang} is introducing the
unitary symmetry breaking terms in some definite way

\begin{eqnarray}
\mu(p)=F_{N}+\frac{1}{3}D_{N}-\frac{1}{3} \beta_{1} \qquad
\mu(n)=-\frac{2}{3}D_{N}-\frac{1}{3} \beta_{1}
\nonumber\\
\mu(\Sigma^{+})=F_{\Sigma}+\frac{1}{3}D_{\Sigma}-\frac{1}{3} \beta_{1}
\qquad
\mu(\Sigma^{-})=-F_{\Sigma}+\frac{1}{3}D_{\Sigma}-\frac{1}{3} \beta_{1}
\nonumber\\
\mu(\Xi^{0})=-\frac{2}{3}D_{\Xi}-\frac{1}{3} \beta_{1} \qquad
\mu(\Xi^{-})=-F_{\Xi}+\frac{1}{3}D_{\Xi}-\frac{1}{3} \beta_{1}
\nonumber\\
\mu(\Lambda^{0})=-\frac{1}{3}D_{\Lambda}-\frac{1}{3} \beta_{1}
\end{eqnarray}

where

$F_{N}=b_{2}+\alpha_{1}+\alpha_{2},\qquad F_{\Sigma}=b_{2}, \qquad
F_{\Xi}=b_{2}-\alpha_{1}+\alpha_{2},$

$D_{N}=b_{1}+\alpha_{3}+\alpha_{4}, \qquad D_{\Sigma}=b_{1}, \qquad
D_{\Xi}=b_{1}-\alpha_{3}+\alpha_{4}, \qquad D_{\Lambda}=
b_{1}-\frac{8}{3}\alpha_{4}$.

The main difference from  other models introducing  symmetry breaking
mechanisms of various kinds consists in the unity operator term
which explicitly breaks the octet form of the electromagnetic current
in $SU(3)_{f}$. Renormalization of the constants $F$ and $D$
can be related to the so-called middle-strong interaction contribution,
which follows from the fact that for the baryons B(qq,q') 
it can be reduced to the form characteristic of mass breaking
terms of the unitary symmetry

\begin{eqnarray}
\mu(p)=F+\frac{1}{3}D+g_{1}+\beta +\pi_{N} \nonumber\\
\mu(n)=-\frac{2}{3}D+g_{1}+\beta -\pi_{N}\nonumber\\
\mu(\Sigma^{+})=F+\frac{1}{3}D+\beta  \nonumber\\
\mu(\Sigma^{-})=-F+\frac{1}{3}D+\beta  \nonumber\\
\mu(\Xi^{0})=-\frac{2}{3}D+g_{2}+\beta \nonumber\\
\mu(\Xi^{-})=-F+\frac{1}{3}D+g_{2}+\beta 
\label{fdmass}
\end{eqnarray}

$F=b_{2} ,\quad
D=b_{1}+\alpha_{1}-\alpha_{2}-\alpha_{3}+\alpha_{4}$,

$g_{1}=\alpha_{1}-\frac{2}{3}\alpha_{3}+
\frac{1}{3}\alpha_{4},\quad
g_{2}=\alpha_{1}-\alpha_{2}-\frac{1}{3}\alpha_{3}+
\frac{1}{3}\alpha_{4}$,

$\beta=-\frac{1}{3}\beta_{1}+\frac{1}{3}(-\alpha_{1}+
\alpha_{2}+\alpha_{3}-\alpha_{4}),
\quad \pi_{N}=\alpha_{2}+\alpha_{3}$.

The results of Eq.(\ref{fdmass}) can be obtained from the following
electromagnetic current (we disregard space-time indices):

\begin{eqnarray}
J^{e-m,symm1}=-F(\overline{B}^{\gamma}_{1}B_{\gamma}^{1}-
\overline{B}^{1}_{\gamma}B_{1}^{\gamma})+
D(\overline{B}^{\gamma}_{1}B_{\gamma}^{1}+
\overline{B}^{1}_{\gamma}B_{1}^{\gamma})+\nonumber\\
g_{1}\overline{B}_{3}^{\gamma}B^{3}_{\gamma}+
g_{2}\overline{B}_{\gamma}^{3}B^{\gamma}_{3}+(\beta-
\frac{2}{3}D)Sp(\overline{B}^{\gamma}_{\beta}B_{\gamma}^{\beta})+
\pi_{N}(\overline{B}_{3}^{1}B_{1}^{3}-\overline{B}_{3}^{2}B_{2}^{3})
\label{fdch}
\end{eqnarray}
Here  $B^{\gamma}_{\eta}$ is a baryon octet, $B^{3}_{1}=p$,
$B^{2}_{3}=\Xi^{0}$ etc.
For the magnetic moment of  the $\Lambda$ hyperon this current gives:
\begin{equation}
\mu(\Lambda)^{symm1}=-\frac{1}{3}b_{1}+
\frac{2}{3}\alpha_{1}-\frac{2}{9}\alpha_{4}-\frac{1}{3}\beta_{1},
\end{equation}
so that
\begin{equation}
\mu(\Lambda)^{symm1}-\mu(\Lambda)^{ChPT}=\frac{2}{3}(\alpha_{1}+\alpha_{4}).
\end{equation}

For $\pi_{N}=0$ the current (\ref{fdch}) yields a direct sum of the
traditional electromagnetic current of the theory of
unitary symmetry \cite{Gl} and the traditional baryon current leading
to the Gell-Mann-Okubo mass relation \cite{GMO}.The parameter $\pi_{N}$
defines the value of contribution of the pion current term 
and is  characteristic of many versions of
the chiral models. Note that with
$\alpha_{1}=-\alpha_{4}$ the ChPT model results reduce to that given
by the current of Eq.(\ref{fdch}). As in \cite{Chang}
$\alpha_{1}=0,32,\alpha_{4}=-0,31$(in $GeV^{-1}$), it goes out that
the chiral model ChPT \cite{Chang} results are in fact given 
by the phenomenological unitary current (\ref{fdch}).

\section{Quark soliton model $\chi QSM $ and unitary symmetry}

$\quad$ In \cite{NJL},\cite{Kim} magnetic moments of baryons were studied 
within the chiral quark soliton model. In this model, known also as
the semibosonized Nambu-Jona-Lasinio model, the baryon can be
considered as $N_{c}$ valence quarks coupled to the polarized Dirac sea
bound by a nontrivial chiral background hedgehog field in the Hartree-Fock
approximation \cite{Kim}. Magnetic moments of baryons were written in the
form \cite{Kim}

\begin{equation}
\left(\begin{array}{ccccccc}\mu(p)\\ \mu(n)\\ \mu(\Lambda)\\
\mu(\Sigma^{+})\\ \mu(\Sigma^{-})\\ \mu(\Xi^{0})\\ \mu(\Xi{-})
\end{array}\right)=
\left(\begin{array}{ccccccc}-8&4&-8&-5&-1&0&8\\6&2&14&5&1&2&4\\
3&1&-9&0&0&0&9\\-8&4&-4&-1&1&0&4\\2&-6&14&5&-1&2&4\\
6&2&-4&-1&-1&0&4\\2&-6&-8&-5&1&0&8\end{array}\right)
\left(\begin{array}{ccccccc}v\\w\\x\\y\\z\\p\\q\end{array}\right)
\label{kim}
\end{equation}
Here the parameters  $v$ and $w$ are linearly related with the usual F,D
coupling constants of the unitary symmetry approach. The parameters
$x,y,z,p,q \simeq m_{s}$ are specific for the model. Upon algebraic
transformations the expressions for 6 baryons $B(qq,q^{'})$
can be rewritten as

\begin{eqnarray}
\mu(p)=F+\frac{1}{3}D-f_{1}+T-3z\nonumber\\
\mu(n)=-\frac{2}{3}D-f_{1}+T+3z\nonumber\\
\mu(\Sigma^{+})=F+\frac{1}{3}D+T \nonumber\\
\mu(\Sigma^{-})=-F+\frac{1}{3}D+T \nonumber\\
\mu(\Xi^{0})=-\frac{2}{3}D-f_{2}+T \nonumber\\
\mu(\Xi^{-})=-F+\frac{1}{3}D-f_{2}+T
\label{alkim}
\end{eqnarray}
where
\begin{eqnarray}
F=-5v+5w-(9x+3y+p)+z\nonumber\\
D=-9v-3w-(13x+7y-4q+p)+3z\nonumber\\
f_{1}=4x+4y-4q-z\nonumber\\
f_{2}=22x+10y-4q+2p-2z\nonumber\\
T=\frac{1}{3}(28x+13y+8q+4p)-z \qquad.
\end{eqnarray}
One can see that the algebraic structures of Eq.(\ref{alkim})
and Eq.(\ref{fdmass}) are the same. It means that magnetic moments
of the octet baryons in the models $\chi QSM$\cite{Kim} and ChPT \cite{Chang}
can be obtained from the unitary electromagnetic
current of the form (we disregard space-time indices)

\begin{eqnarray}
J^{e-m,symm2}=-F(\overline{B}^{\gamma}_{1}B_{\gamma}^{1}-
\overline{B}^{1}_{\gamma}B_{1}^{\gamma})+
D(\overline{B}^{\gamma}_{1}B_{\gamma}^{1}+
\overline{B}^{1}_{\gamma}B_{1}^{\gamma})- \nonumber\\
f_{1}\overline{B}_{3}^{\gamma}B^{3}_{\gamma}-
f_{2}\overline{B}_{\gamma}^{3}B^{\gamma}_{3}+(T-
\frac{2}{3}D)Sp(\overline{B}^{\gamma}_{\beta}B_{\gamma}^{\beta})+
3z(\overline{B}_{3}^{2}B_{2}^{3}-\overline{B}_{3}^{1}B_{1}^{3})
\label{fdkim}
\end{eqnarray}
With this current the magnetic moment of the $\Lambda$ hyperon reads:

\begin{equation}
\mu(\Lambda)^{symm2}=-\frac{1}{3}D-(8x+5y-8q-z)
\end{equation}
which differs from that given by Eq.(\ref{kim}) 

\begin{equation}
\mu(\Lambda)^{symm2}-\mu(\Lambda)^{\chi QSM}=\frac{1}{3}
(16x-8y-7q+p)
\end{equation}

Eqs.(\ref{kim}) and (\ref{chang}) reduce to each other through 
the relations between the parameters

\begin{eqnarray}
b_{1}=-(9v+3w)-\frac{1}{2}(42x+6y-15q+3p),\nonumber\\
b_{2}=-5(v-w)-(9x+3y+-z+p)\nonumber\\
\alpha_{1}=-\frac{1}{4}(94x+34y-31q+7p), \quad
\alpha_{2}=\frac{3}{2}(9x+3y-z+p),\nonumber\\
\alpha_{3}=-\frac{3}{2}(9x+3y+z+p),\quad
\alpha_{4}=\frac{9}{4}(14x+2y-5q+p),\nonumber\\
\beta_{1}=-\frac{9}{2}(8x+2y+q+p) \qquad \qquad,
\label{coin}
\end{eqnarray}
where now

\begin{equation}
9 \alpha_{1}+15 \alpha_{2}-15 \alpha_{3}+3 \alpha_{4}+8 \beta_{1}=0.
\label{beta}
\end{equation}
These formulae yield the following relation between the
octet baryon magnetic moments derived in \cite{Hong} and \cite{Kim}

\begin{equation}
-12\mu(p)-7\mu(n)+7\mu(\Sigma^{-})+
22\mu(\Sigma^{+})-12\mu(\Lambda^{0})+3\mu(\Xi^{-})+
23\mu(\Xi^{0})=0 \quad .
\end{equation}

The relations (\ref{coin}) and (\ref{beta}) close our proof of the
practical coincidence of the magnetic
moment  description in the framework of the ChPT \cite{Chang} model
and the $\chi QSM $ \cite{Kim} one.

\section{Summary and conclusion}

It has been shown that the algebraic schemes of the models
\cite{Chang} and \cite{Kim} for the predictions of the octet
baryon magnetic moments have proved to be practically identical. Moreover,
the expressions for the magnetic moments B(qq,q') in these 
models are those given by the unitary model with the
phenomenological electromagnetic current given by Eq.(\ref{fdch}) or
Eq.(\ref{fdkim}). The main difference of the models \cite{Chang}
and \cite{Kim} from a direct sum of the traditional unitary
electromagnetic and middle-strong baryon currents lies  in the
terms due to pion current contribution which are written 
excplicitly in Eqs.(\ref{fdch}) and (\ref{fdkim}). The only real
difference between our phenomenological current predictions and those of
\cite{Chang} and \cite{Kim} is in the formula for the 
magnetic moment of the hyperon $\Lambda$. 
This difference may prove to have
deeper meaning as $\Lambda$-hyperon being composed of all
different quarks is characterized by zero values of isotopic
spin and hypercharge. Quantitavely it occurs not to be very 
important as due to approximate equality $\alpha_{1}=-\alpha_{4}$
the $\Lambda$ magnetic moment proves to be practically the same
in the phenomenological model given by Eq.(\ref{fdch}) and ChPT
\cite{Chang}.

In general, the analysis of the baryon magnetic moments in the
framework of these models has shown once more that unitary
symmetry is the basis which could be hidden in any dynamical
model pretending to an adequate description of the electromagnetic
properties of baryons.

\end{document}